\documentclass[journal]{journal}

\ifCLASSINFOpdf
    \usepackage{xcolor}
    \usepackage{graphicx}
    \usepackage{multirow}
    \usepackage{array}
    \usepackage[a4paper,margin=1in]{geometry}
    \usepackage{graphicx} 
    \usepackage{array} 
    \usepackage{booktabs}
\else
\fi

\begin{document}

\title{Multilingual Email Phishing Attacks Detection using OSINT and Machine Learning}

\author{Panharith An, Rana Shafi, Tionge Mughogho, Onyango Allan  Onyango}

\markboth{Journal of \LaTeX\ Class Files,~Vol.~6, No.~1, January~2007}%
{Shell \MakeLowercase{\textit{et al.}}: Bare Demo of IEEEtran.cls for Journals}

\maketitle
\thispagestyle{empty}

\begin{abstract}
Email phishing remains a prevalent cyber threat, targeting victims to extract sensitive information or deploy malicious software. This paper explores the integration of open-source intelligence (OSINT) tools and machine learning (ML) models to enhance phishing detection across multilingual datasets. Using Nmap and theHarvester, this study extracted 17 features, including domain names, IP addresses, and open ports, to improve detection accuracy. Multilingual email datasets, including English and Arabic, were analyzed to address the limitations of ML models trained predominantly on English data. Experiments with five classification algorithms: Decision Tree, Random Forest, Support Vector Machine, XGBoost, and Multinomial Naïve Bayes. It revealed that Random Forest achieved the highest performance, with an accuracy of 97.37\% for both English and Arabic datasets. For OSINT-enhanced datasets, the model demonstrated an improvement in accuracy compared to baseline models without OSINT features. These findings highlight the potential of combining OSINT tools with advanced ML models to detect phishing emails more effectively across diverse languages and contexts. This study contributes an approach to phishing detection by incorporating OSINT features and evaluating their impact on multilingual datasets, addressing a critical gap in cybersecurity research.

\end{abstract}

\begin{IEEEkeywords}
Email Phishing, Open-source Intelligence, OSINT, Multi-lingual, Multilingual, Machine Learning, Classification Algorithms, Artificial Intelligence
\end{IEEEkeywords}

\IEEEpeerreviewmaketitle

\section{Introduction}
\IEEEPARstart{E}{mail} phishing is a malicious practice wherein perpetrators dispatch fraudulent messages to victims to manipulate them into revealing private details or running detrimental software. Phishing has become one of the most common social engineering attacks targeting users’ emails to fraudulently steal confidential and sensitive information. Phishing attacks have grown
to be one of the leading paths to bigger attacks such as ransomware attacks and denial of service attacks \cite{orunsolu2019predictive}. Post-COVID-19, the world has faced an increase in online activities such as online shopping, online banking, and digital banking \cite{de2020impact}. This has undoubtedly increased the attack surface and vectors for these phishing attacks \cite{shombot2024application}.

The issue of multi-lingual phishing can also not be overlooked, as most ML models feed on phishing indicators in the English language, hence limiting their detectability of phishing attacks in other languages. \cite{fette2007learning} identified eleven parameters deemed essential for effective fraud filters in their investigation of email phishing detection. However, gaps remain in addressing challenges such as the ability to detect phishing emails in multiple languages, the lack of generalized models that adapt across different-language datasets, and OSINT-enhanced features that will improve the models' accuracy. Open-source Intelligence (OSINT) refers to a sector of intelligence that gathers and examines facts exclusively from freely and publicly accessible sources \cite{nordine2019osint}.

To address the gaps, this research looks to leverage these parameters to include indicators across multiple languages. As a result, this research primarily focuses on using various machine learning to develop different models that can detect phishing attacks while accounting for the cross-language gap. 

In a multilingual context, this study would indicate if the models trained by OSINT-enhanced features are more likely to outperform the models trained by the original dataset, which has fewer features.

Key contributions:
\begin{itemize}
    \item A comparison study among four different datasets and five different machine-learning classification algorithms to detect phishing emails in both English and Arabic.
    \item A study comparing the performance of ML models on datasets with and without OSINT-driven feature selection to evaluate its impact on improving detection accuracy.
\end{itemize}

\section{Research Questions}
\begin{itemize}
    \item What OSINT tools can be used to extract features from our English and Arabic datasets?
    \item Which ML models can be used to train the datasets in detecting phishing emails?
    \item Can the OSINT-enhanced datasets in both languages be more accurately detected by the same algorithm compared to the original dataset?
\end{itemize}

\section{Literature Review}

\subsection{Search Process}
This study aims to focus on two different topics: OSINT and machine learning applied to email phishing detection. For better understanding, a systematic literature review (SLR) was adopted across four databases: Scopus, Web of Science, IEEE Xplore, and Google Scholar. Query strings:
\begin{itemize}
    \item (“Machine Learning” OR “Artificial Intelligence” OR “ML” OR “AI”) AND (“Phishing” OR “Phishing Attacks”  OR “Cybersecurity” OR “Cyber Security”)
    \item (“Machine Learning” OR “Artificial Intelligence” OR “ML” OR “AI”) AND (“OSINT” OR “Open-source Intelligence” OR “Open source intelligence”)
    \item (“OSINT” OR “Open-source Intelligence” OR “Open source intelligence”) AND (“Phishing” OR “Phishing Attacks” OR “Cybersecurity” OR “Cyber Security”)
\end{itemize}

\begin{figure}[h]
    \centering
    \includegraphics[width=0.48\textwidth]{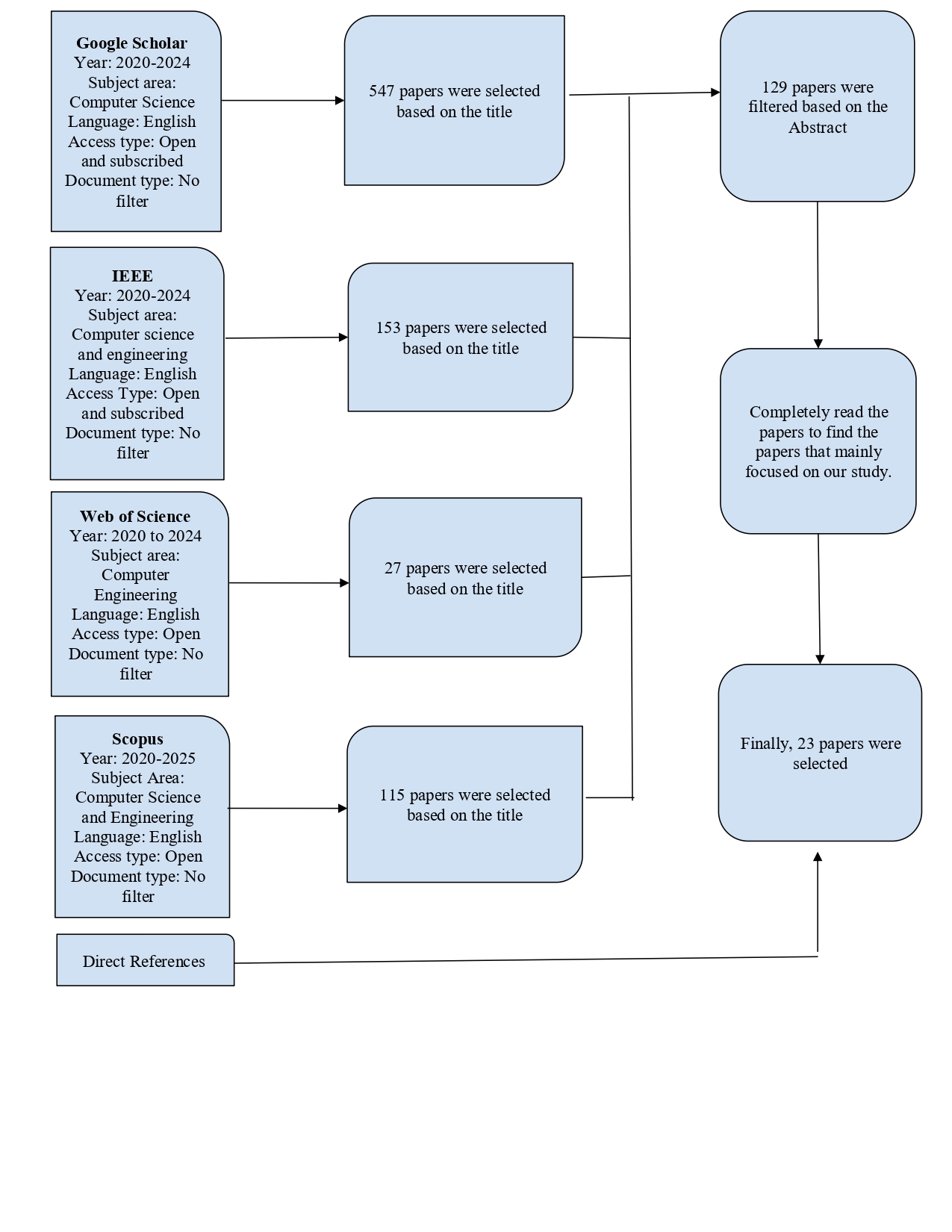}
    \caption{Search Process Diagram in SLR}
    \label{fig:Picture8}
\end{figure}

Figure~\ref{fig:Picture8} shows the process of our search strategies and results.

Table \ref{tab:slr_review} summarizes related works in the domains of OSINT investigation, phishing attacks, and machine learning. While these studies provide valuable insights and advancements in phishing detection, significant gaps remain. These include limitations in multilingual phishing detection, inadequate integration of OSINT features, and challenges in feature extraction optimization. This study seeks to address these gaps by combining OSINT tools with advanced machine learning algorithms to enhance phishing detection accuracy across diverse languages and contexts.

\section{Research Method}
\subsection{Data Collection}
\begin{figure}
    \centering
    \includegraphics[width=0.5\linewidth]{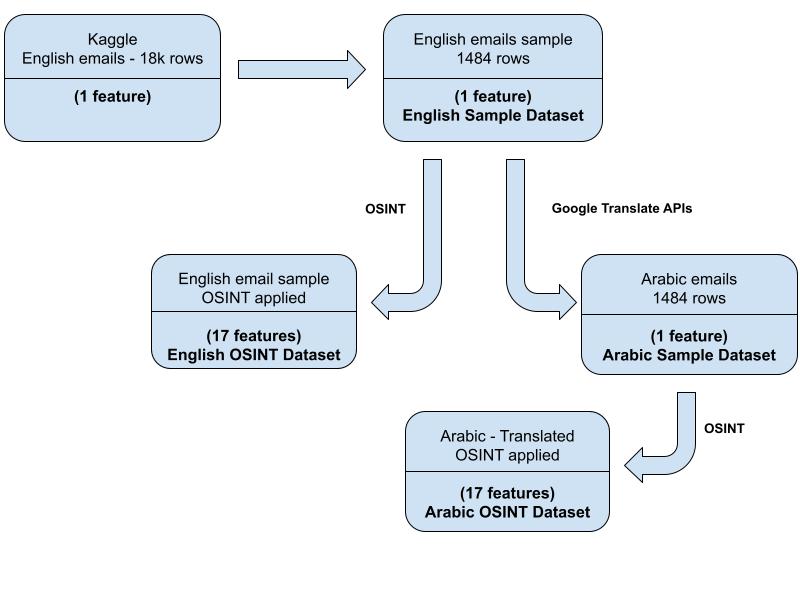}
    \caption{Choosing and generating the dataset groups}
    \label{fig:datasets}
\end{figure}

This study will experiment with \textcolor{red}{} different groups of datasets. Figure \ref{fig:datasets} depicts how each group of the dataset was chosen or generated.
\begin{itemize}
    \item group 1 (English): English phishing emails from Kaggle \cite{kaggle_phishing}
    \item group 2 (English Sample): randomly sampling 1,484 rows from group 1.
    \item group 3 (English OSINT): obtained by adding 17 OSINT features to group 2. 
    \item group 4 (Arabic sample): obtained using Google Translate API on group 2, and verified by a native Arabic speaker (over 90\% correctness)
    \item group 5 (Arabic OSINT): obtained by adding 17 OSINT features to group 4.
\end{itemize}

\subsubsection{Choosing the Languages}
It is notable that Arabic was chosen as another language for this multilingual study because the Arabic language differs dramatically from English, which uses Roman alphabet. This study seeks to compare the performance of ML models in the two distinct languages.

The English dataset, published in 2023, originally contains one feature (Email Text) and one label (Email Type: Phishing or Safe). The number of rows originally is 18,651. Using SHA-256 hashing to remove the duplicates, the dataset is reduced to 17,522 non-duplicates (6,544 records are in phishing type). With its diversity and size, this 2023-published dataset was also studied by other researchers such as \cite{kaggle_paper1}, \cite{kaggle_paper2}, and \cite{kaggle_paper3}, highlighting its usefulness and contributions to the scientific community.

From the above-mentioned English dataset, this study randomly sampled 1,484 rows that contained proper URLs, domain names, and emails. We named these samples the English Sample dataset, from which we extracted OSINT features and generated the English OSINT dataset. 

It is also important to note that regardless of the small size of the OSINT dataset groups, this study is based on the principles of random sampling, ensuring that each data point had an equal chance of being selected, thereby preserving the representativeness of the dataset. This approach minimizes selection bias and maintains the statistical integrity of the sample, making the findings applicable to the broader dataset of 17 thousand rows.
\clearpage  
\begin{table}[h!]
\centering
\renewcommand{\arraystretch}{1.5} 
\begin{tabular}{|p{3cm}|p{2cm}|p{2cm}|p{4.5cm}|p{4cm}|} 
\hline
\textbf{Reference} & \textbf{Method} & \textbf{Dataset} & \textbf{Descriptions} & \textbf{Limitations} \\ \hline
A Feature Extraction Approach for the Detection of Phishing Websites Using Machine Learning \cite{sricharangundla_2023_a}& DT, RF, Multilayer perceptrons, XGBoost, SVM, KNN, and NB & Phishing URLs. The source is not mentioned. & ML techniques were analyzed and implemented to accurately detect phishing attacks with less time complexity. The MLPs technique provided the best precision score and the highest NDCG score. & The feature extraction algorithms needed to be more optimized, and new features needed to be chosen carefully to increase accuracy. The dataset was tested with basic classification algorithms, which led to lower accuracy than other algorithms. \\ \hline
Phishing Attacks Facilitated by Open-Source Intelligence \cite{maryam_2023}& OSINT (Maltego) & Gathering public email addresses from social media platforms. & Phishing attacks were launched using OSINT to test the response of people to spoofed emails. 5 out of the 20 participants fell for the phishing attacks. & The sample size of 20 participants which may potentially be non-diverse could affect the generalizability of the findings and fail to account for variability in user behavior across different demographics. \\ \hline
Multi-Language Spam/Phishing Classification by Email Body Text: Toward Automated Security Incident Investigation \cite{rastenis_2021}& SVM, RF, DT, NB, LR, KNN & Nazario (phishing emails), SpamAssassin (spam emails), and a combined dataset from Vilnius Gediminas Technical University. & Multiple ML models were used to measure their accuracy in identifying multilingual phishing emails. The support vector machine had the highest accuracy (84.0\% ± 1.6\%), however, was one of the slowest solutions. & Deeper spam/phishing email classification performance analysis could be executed to increase performance by adapting feature optimization (including header and formatting-related features, etc.), and evaluating deep-learning solution suitability for this task. \\ \hline
A Comparison of Machine Learning Algorithms for Multilingual Phishing Detection \cite{staples_2023}& XML Roberta, LR, SVM, RF, GPT-4, and GPT-3 & English, French, and Russian spam emails from Enron. & An accuracy comparison of ML models in detecting multilingual phishing emails. XLM-Roberta performs the best out of all of the tested models in terms of accuracy. & The dataset used as French and Russian dataset was translated from English; therefore, it may not fully capture linguistic nuances and cultural context, potentially leading to false positives or negatives. \\ \hline
Automation of the Information Collection Process by OSINT Methods for Penetration Testing During Information Security Audit \cite{bryushinin_2022}& Programmatic data collection with OSINT & Extracting emails, phone numbers, organization addresses, etc., from the Internet. & The use of OSINT methods to automate data collection from open sources for penetration testing. & Software developed does not describe the level of sub-pages it could dive into. No clear relation between software development and pen-testing or information security audit exists. \\ \hline
Machine Learning Techniques for Detecting Phishing URL Attacks \cite{mosa_2023}& Neural Networks (NN), Naïve Bayes, and Adaboost. & Over 11k Websites from Kaggle. & The confusion matrices of three ML models in detecting phishing URL attacks were studied and their accuracies were observed. Neural Networks (NN), Naïve Bayes, and Adaboost were studied, and the results indicated that the accuracies achieved were 90.23\%, 92.97\%, and 95.43\%, respectively. & The model is indifferent whether the website’s URL is active or contains an error. Short links, sensitive phrases, and phishing URLs that do not replicate other websites will be misclassified by the system. \\ \hline
An application for predicting phishing attacks: A case of implementing a support vector machine learning model \cite{emmanuel_2024}& SVM & Website Phishing URLs from the work of its reference study. & A GUI was also created for the user to report any email that turns out to be a phishing email. The outcome of the model shows that the polynomial function performed better with 84.5\% accuracy, while the radial basis function had an accuracy score of 82.6\%. & The SVM prediction model is based on a dataset of about 1,400 records, which may have had an impact on the model's accuracy. \\ \hline
Detecting Phishing Domains Using Machine Learning \cite{alnemari_2023}& ANN, SVM, DTs, and RF algorithms. & URLs and UCI phishing domains dataset. & Developed and compared four models for investigating the efficiency of using machine learning to detect phishing domains. The findings show that the model based on the random forest technique is the most accurate of the other four techniques and outperforms other solutions in the literature. & Future work includes examining more machine learning algorithm techniques for phishing domains. \\ \hline
Phishing Website Detection through Multi-Model Analysis of HTML Content \cite{fukan_2024_phishing} & MLP and NLP models & Phishing URLs from OpenPhish. & MLP model and NLP models were used to create an advanced phishing detection model that focuses on HTML content. The standalone MLP model achieved an accuracy of 89.92\%. The NLP-1 and NLP-2 models achieved accuracies of 93.84\% and 96.76\%, respectively. The MultiText-LP model, a fusion of NLP-1, NLP-2, and MLP models achieved an accuracy of 97.18\%, illustrating the synergistic effect of combining both approaches. & Using two pre-trained and one MLP model simultaneously requires a powerful GPU like A4000. Standard GPUs may struggle with their size, making training and deployment less efficient, especially in places with limited resources. \\ \hline
\end{tabular}
\caption{Summary of Literature Reviews and Related Works}
\label{tab:slr_review}
\end{table}
\clearpage

Additionally, random sampling is a widely accepted method in research to manage large datasets effectively while maintaining reliable and valid results.

After that, this study translated the English-OSINT dataset into Arabic, which generated our two last datasets: the Arabic Sample dataset and the Arabic OSINT dataset.

\subsubsection{Preprocessing}
The next step is pre-processing the dataset to prepare it for machine learning to ensure the dataset is structured, reliable, and ready for training, improving model performance. This includes cleaning to remove noise and duplicates, handling missing values through imputation or default categories, and encoding categorical features using one-hot or label encoding.

\subsubsection{Balancing the dataset}
Balancing the dataset is important to prevent bias in the machine learning model and ensure fair representation of all classes during training. After balancing the dataset using the undersampling technique, 504 emails remain: 252 phishing emails and 252 safe emails. 

\subsubsection{OSINT Feature Collection}
\begin{figure}[h]
    \centering
    \includegraphics[width=8cm, height=4cm]{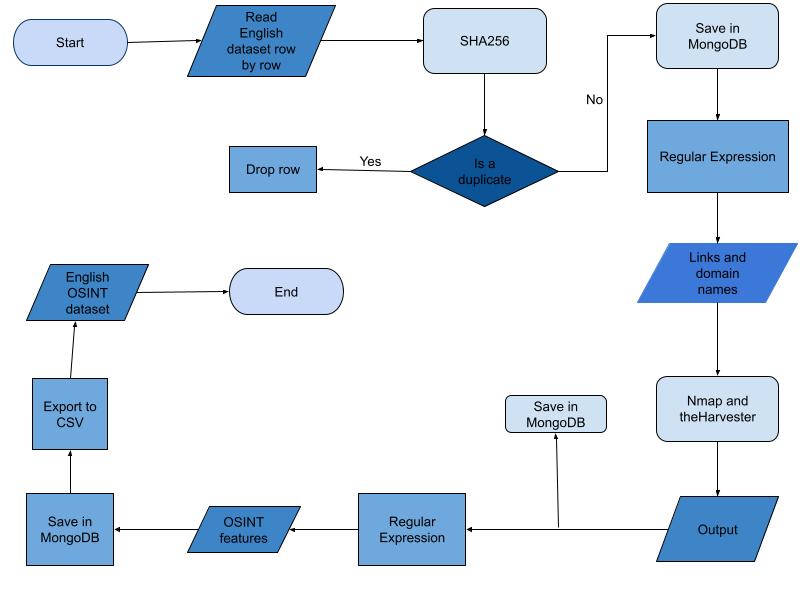}
    \caption{Flowchart of OSINT Data Collection \& Feature Extraction}
    \label{fig:flowchart-osint-feat}
\end{figure}

This study developed a Python program to automate raw output, with the extracted domain names and URLs, using two OSINT tools: Nmap and theHarvester. Figure \ref{fig:flowchart-osint-feat} depicts the process of OSINT feature extraction using Python. 

The Python program is based on the following commands, which produce 17 features shown in Table \ref{tab:osint_feat}:
\begin{itemize}
    \item \texttt{nmap -Pn -T4 --max-retries 3 [domain name]} \cite{nmap_ip}, \cite{nmap_max}
    
    \item \texttt{python theHarvester.py -d [domain name] -l 500 -b all} \cite{samrudhamhatre_2024_osintbased}
\end{itemize}

\begin{table}[!t]
\centering
\begin{tabular}{|p{2.4cm}|p{2.1cm}|p{2.1cm}|}
\hline
\textbf{Feature} & \textbf{Description} & \textbf{Example} \\ \hline
hostname & Domain name retrieved from the email server. & \texttt{\scriptsize www.oreilly.com}\\ \hline
host\_up & Host device reachability (1 = up, 0 = down). & \texttt{1, 0} (Sum will be calculated)\\ \hline
alternate\_ip\_count & Count of alternative IP addresses.& \texttt{2, 6} (Sum will be calculated)\\ \hline
ip\_address & Main IP mapped to the domain name. & \texttt{\scriptsize 34.169.83.167, 23.52.38.113} \\ \hline
\scriptsize common\_web\_ports\_open & Open web server ports (0 = closed, 1 = open). & \texttt{0, 1} (Sum will be calculated)\\ \hline
open\_ports\_count & Total open TCP/UDP ports.& \texttt{4, 8} (Sum will be calculated)\\ \hline
filtered\_ports\_count & Count of filtered TCP/UDP ports. & \texttt{2, 1} (Sum will be calculated)\\ \hline
open\_ports & Total listening ports on the device. & \texttt{80, 443} \\ \hline
rdns\_record & Reverse DNS record for the IP. & \texttt{example.com} \\ \hline
https\_supported & HTTPS support (0 = no, 1 = yes).& \texttt{1, 0} (Sum will be calculated for each URL)\\ \hline
services & Running services on open ports. & \texttt{HTTPS, FTP, mysql, ssh, pop3} \\ \hline
host\_found & Host record detected.& \texttt{1, 0} (Sum will be calculated)\\ \hline
interesting\_url & URL contains suspicious content. & \texttt{3, 10} (Sum will be calculated) \\ \hline
asn\_found & Autonomous system number identified for the IP.& \texttt{0, 3, 11} (Sum will be calculated)\\ \hline
ip\_found & Email contains an IP address. & \texttt{1, 3} (Sum will be calculated)\\ \hline
latency & Response delay during the scan (in seconds).& \texttt{0.037, 0.098, 0.100} (Sum will be calculated)\\ \hline
scan\_duration & Total time taken to complete the scan (in seconds).& \texttt{43.64, 55.99} (Sum will be calculated)\\ \hline
\end{tabular}
\caption{Features and Descriptions from OSINT}
\label{tab:osint_feat}
\end{table}

\subsection{System Model in the Experiment}

This study conducted 20 experiments with the four different datasets mentioned above, using five different machine-learning classification algorithms: Decision Tree (DT), Random Forest (RF), Support Vector Machine (SVM), Multinomial Naive Bayes (MultinomialNB), and XGBoost. These algorithms were chosen in this experiment because of their proven effectiveness in handling classification tasks, their ability to manage diverse feature sets, and their widespread application in phishing detection and OSINT-related studies. Additionally, these algorithms were also experimented with in the related works mentioned in Table \ref{tab:slr_review}, providing a baseline for comparison and validation of this study’s methodology and results.

We utilized Grid Search to find the best parameters for our models. Doing so ensures each classification model has the best performance result for this comparative study.

\begin{figure}[h]
    \centering
    \includegraphics[width=8cm, height=4cm]{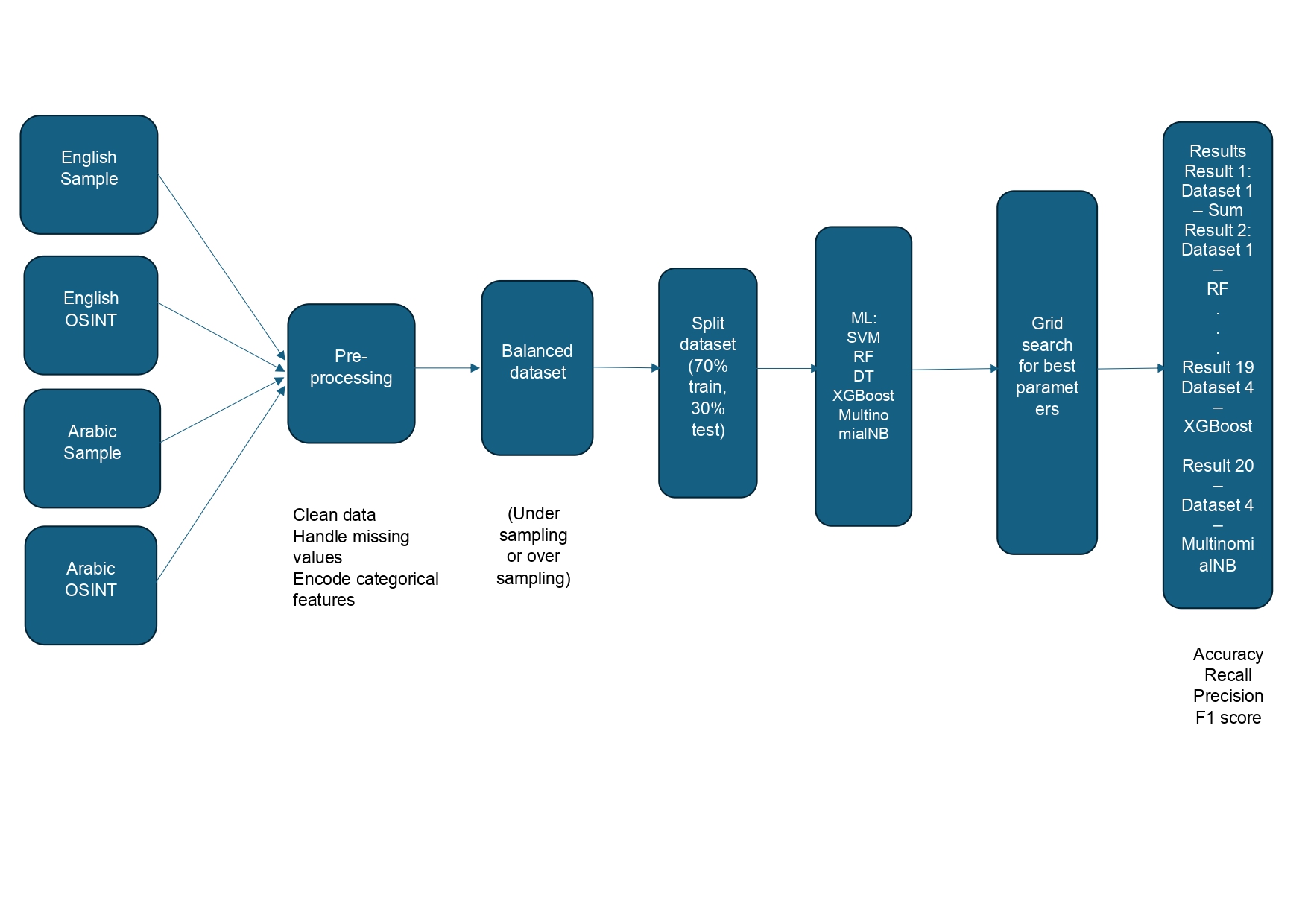}
    \caption{Diagram of Model Trainings for Email Phishing Detection}
    \label{fig:training}
\end{figure}

Figure \ref{fig:training} summarizes the training process of ML models for email phishing detection in this experiment.  

\section{Results}

This section of the paper presents our experimental setting and findings. We provide a detailed analysis of our results, comparing each model's performance on the datasets before and after adding our OSINT features.
\subsection{Experimental Setting}
Our experiments were conducted in VSCode, using Jupyter kernel. We used 4 Python libraries: Scikit-learn, Pandas, Numpy, and Matplotlib. The latter was used to visualize our dataset's size before and after balancing, as well as the confusion matrices after each test. Table \ref{tab:ml-params} lists the hyperparameters used for training the models.
\begin{table}[htbp]
\centering
\caption{Machine Learning Models Hyperparameters after Finetuning}
\begin{tabular}{|l|l|l|}
\hline
\textbf{Model} & \textbf{Hyper-parameter} & \textbf{Value} \\
\hline
\multirow{4}{*}{Decision Tree}
& criterion & ``entropy'' \\
& max\_depth & None \\
& min\_samples\_split & 5 \\
& min\_samples\_leaf & 1 \\
\hline
\multirow{4}{*}{Random Forest} 
& n\_estimators & 100 \\
& max\_depth & None \\
& min\_samples\_split & 2 \\
& min\_samples\_leaf & 1 \\
\hline
\multirow{3}{*}{SVM} 
& C & 100 \\
& kernel & ``rbf'' \\
& gamma & 0.1 \\
\hline
\multirow{3}{*}{XGBoost}  
& n\_estimators & 100 \\
& eval\_metric & ``logloss'' \\
& use\_label\_encoder & False \\
\hline
MultinomialNB & alpha & 0.1 \\
\hline
\end{tabular}
\label{tab:ml-params}
\end{table}

\subsection{Experimental Results}
To evaluate the performance of our models, we focused on key metrics such as accuracy, F1 score, recall, and precision. 
We also included the confusion matrix of each test.
Table \ref{tab:classifier_comparison1} provides the experimental results obtained after training each model on both the English sample dataset and the English OSINT dataset. 
\begin{table}[h!]
\caption{Experimental Results on English dataset before and after adding OSINT features (in \%)}
\setlength{\tabcolsep}{2pt}
\begin{tabular}{|l|l|l|l|l|l|}
\hline
\textbf{Classifier} & \textbf{Dataset} & \textbf{Accuracy} & \textbf{F1 Score} & \textbf{Precision} & \textbf{Recall}
\\ 
\hline
DT   & English Sample   & 83.55      & 83.44      & 84.00       & 82.89             
\\  
                & English OSINT    & 86.84      & 86.30      & 90.00       & 82.89              
\\ \hline
RF   & English Sample   & 96.71      & 96.77      & 94.94       & 98.68              
\\ 
                & English OSINT    & 97.37      & 97.30      & 100       & 94.74      
\\ \hline
SVM             & English Sample   & 84.21      & 85.71      & 78.26       & 94.74                
\\ 
                & English OSINT    & 88.16      & 87.84     & 90.28       & 85.53  
\\ \hline
XGBoost         & English Sample   & 95.39      & 95.54      & 92.59       & 98.68                
\\ 
                & English OSINT    & 96.71      & 96.73      & 96.10       & 97.37 
\\ 
\hline
MNB     & English Sample   & 96.71      & 96.71      & 96.72       & 96.71             \\ 
                & English OSINT    & 96.05      & 96.10      & 94.87       & 97.37    
\\ 
\hline
\end{tabular}

\label{tab:classifier_comparison1}
\end{table}

Starting with Decision Tree, which demonstrated improvements after adding OSINT features to our dataset, increasing its accuracy from 83.55\% to 86.84\%. F1 score and precision increased similarly, while recall remained constant.
\\
\\Random Forest performed exceptionally well in both datasets, improving its accuracy from 96.71\% to 97.37\%. It exhibited a perfect precision score and an increased f1 score with our OSINT dataset. However, it showed a decrease in recall from 98.68\% to 94.74\%.
\\
\\SVM showed significant improvement when enhanced with OSINT features. Its accuracy increased from 84.21\% to 88.16\%, while precision saw a significant increase from 78.26\% to 90.28\%. However, it showed a decline in recall, decreasing from 94.74\% to 85.53\%.
\\
\\Similarly to RF, XGBoost demonstrated a strong performance, with an improved accuracy of 96.71\%. It improved its overall score across all metrics, making it the second-best performer after Random Forest.
\\
\\Multinomial NB is the only classifier that showed a slight decrease in accuracy after OSINT feature enhancement, which dropped from 96.71\% to 96.05\%. The recall, however, increased from 96.71\% to 97.37\%, reflecting the model's improvement in identifying phishing emails.
\\
\\Moving on to Table \ref{tab:classifier_comparison2}, which highlights the experimental results obtained using our translated Arabic datasets. 

\begin{table}[h!]
\centering
\caption{Experimental Results on Arabic dataset before and after adding OSINT features (in \%)}
\setlength{\tabcolsep}{2pt}
\begin{tabular}{|l|l|l|l|l|l|}
\hline
\textbf{Classifier} & \textbf{Dataset} & \textbf{Accuracy} & \textbf{F1} & \textbf{Precision} & \textbf{Recall}
\\
\hline
DT  & Arabic Sample   & 86.84      & 86.84      & 86.84       & 86.84             
\\  
                & Arabic OSINT    & 88.16      & 88.46      & 86.25       & 90.79              
\\ \hline
RF   & Arabic Sample   & 95.39      & 95.17      & 100      & 90.79              
\\ 
                & Arabic OSINT    & 97.37     & 97.33      & 98.65       & 96.05       
\\ \hline
SVM             & Arabic Dataset   & 88.16      &  87.67     & 91.43       &  84.21               
\\ 
                & Arabic OSINT    & 91.45      & 91.50     & 90.91       & 92.11  
\\ \hline
XGBoost         & Arabic Sample   & 95.39      & 95.48      & 93.67       & 97.37                
\\ 
                & Arabic OSINT    & 96.71      & 96.77      & 94.94       & 98.68 
\\ 
\hline
MNB     & Arabic Sample   & 96.05     &  96.15     & 93.75       &  98.68                
\\ 
                & Arabic OSINT    & 93.42      & 93.59      & 91.25       & 96.05    
\\ 
\hline
\end{tabular}
\label{tab:classifier_comparison2}
\end{table}

As observed from a first glance at the table, the performance of each model remains stable throughout this experiment. The 4 models, DT, RF, SVM, and XGBoost, showed improved accuracies from 86.84\% to 88.16\%, 95.39\% to 97.37\%, 88.16\% to 91.45\%, 95.39\% to 96.71\%, respectively. Precision decreased in the first 3 models, except with XGBoost, where it improved from 93.67\% to 94.94\%. Recall and f1 score showed great improvement in the 4 models altogether. 
\\Multinomial NB, similar to its performance on the English dataset, decreased its accuracy from 96.05\% to 93.42\%. Precision, recall, and f1 score decreased similarly. However, the overall scores were still considerably high.
\\
\\We created Table \ref{tab:confusion_comparison1} and Table \ref{tab:confusion_comparison2} to better visualize each model's ability to correctly classify phishing and safe emails.

\begin{table}[h]
\centering
\caption{Comparison of Confusion Matrices between English Sample and English OSINT dataset}
\renewcommand{\arraystretch}{1}
\setlength{\tabcolsep}{6pt}
\begin{tabular}{l|cc|cc}
\toprule
Model & \multicolumn{2}{c|}{English Sample} & \multicolumn{2}{c}{English OSINT}\\
\hline
\midrule
\multirow{2}{*}{DT} 
& 64 & 12  & 69 & 7 \\
& 13  & 63 & 13  & 63 \\
\hline
\midrule
\multirow{2}{*}{RF}
& 72 & 4  & 76 & 0 \\
& 1  & 75 & 4 & 72 \\
\midrule
\multirow{2}{*}{SVM}
& 56 & 20   & 69 & 7 \\
& 4  & 72 & 11 & 65 \\
\hline
\midrule
\multirow{2}{*}{XGBoost}
& 70 & 6  & 73 & 3 \\
& 1  & 75 & 2 & 74 \\
\hline
\midrule
\multirow{2}{*}{MultinomialNB}
& 77 & 0  & 72 & 4 \\
& 5  & 70 & 2 & 74 \\
\bottomrule
\end{tabular}
\label{tab:confusion_comparison1}
\end{table}

\begin{table}[h]
\centering
\caption{Comparison of Confusion Matrices between Arabic Sample and Arabic OSINT dataset}
\renewcommand{\arraystretch}{1}
\setlength{\tabcolsep}{6pt}
\begin{tabular}{l|cc|cc}
\toprule
Model & \multicolumn{2}{c|}{Arabic Sample} & \multicolumn{2}{c}{Arabic OSINT} 
\\
\hline
\midrule
\multirow{2}{*}{DT} 
& 66 & 10  & 65 & 11 \\
& 10  & 66 & 7  & 69 \\
\hline
\midrule
\multirow{2}{*}{RF}
& 69 & 7  & 75 & 1 \\
& 0  & 76 & 3 & 73 \\
\hline
\midrule
\multirow{2}{*}{SVM}
& 64 & 12   & 69 & 7 \\
& 6  & 70 & 6 & 70 \\
\hline
\midrule
\multirow{2}{*}{XGBoost}
& 71 & 5  & 72 & 4 \\
& 2  & 74 & 1 & 75 \\
\hline
\midrule
\multirow{2}{*}{MultinomialNB}
& 71 & 5  & 69 & 7 \\
& 0  & 76 & 3 & 73 \\
\bottomrule
\end{tabular}
\label{tab:confusion_comparison2}
\end{table}

By comparing DT, RF, SVM, and XGBoost's performance with the English Sample dataset before and after adding OSINT features (Table \ref{tab:confusion_comparison1}), we notice that the number of false positives decreased, while the number of true positives increased. However, this is not the case for true negatives and false negatives, which decreased and increased, respectively, across all models except DT, which maintained the same numbers.
\\However, if we examine Multinomial NB's results, we notice that it improved in areas where the other models did not perform well and vice versa. For example, it increased its ability to identify true negatives while at the same decreasing the number of false negatives.

Moving on to Table \ref{tab:confusion_comparison2}, we can draw the same observations about true positives; the number of true positives increased with the first 4 models. However, the number of true negatives only improved with DT and XGBoost, remaining the same with SVM while decreasing with RF. As for false negatives, only DT and XGBoost reduced their occurrences. 
\\Multinomial NB's performance on the Arabic dataset showcased the worst performance, with a decrease in the number of true positives and true negatives and an increase in the number of false positives and false negatives. 

\section{Discussion}
Our experimental results show increased model accuracies after including OSINT features in the dataset. This increase is consistent for both English and Arabic datasets. RF demonstrated the highest accuracy of 97.37\%, followed by XGBoost, which scored an improved accuracy of 96.71\% in both datasets. On the other hand, Multinomial NB was the only model that scored a lower accuracy and performance after training on the OSINT dataset. A possible explanation would be Multinomial NB's inability to handle strongly correlated features, as is the case in our dataset. Our OSINT extracted features are interconnected, and that in itself could have been enough to decline the model's performance.

While the OSINT enhancement generally improved overall accuracy and reduced false positives, its effect on false negatives was mixed. Most models saw an increase in false negatives, mainly SVM in the English dataset. A possible reason for this increase is the use of a small dataset with 19 features, which could affect the models' performance. The more features increase, the larger the dataset needs to be.

The models were efficient in reducing false alarms while maintaining and improving their abilities to detect positive instances. However, their ability to classify false negatives declined.

This paper \cite{sricharangundla_2023_a} experimented with a dataset consisting of 10000 phishing URLs and, similar to our experiment, trained the models using a 70:30 training and testing dataset split. It used 50 features. Their results show that XGBoost demonstrates the best performance.

In this paper \cite{feature-ext2}, the database used contains 11215 
records and 21 features, and was tested using Random Forest, achieving an accuracy of 100\%. 

Rishikesh et al. \cite{feature-ext3}
used a dataset of 36711 URLs from which they extracted 16 features. Their results after splitting the dataset into a 70:30 ratio show a higher performance in Random Forest, which scored an accuracy of 96.84\%.

Returning to our own research, our results align with other research findings in that RF and XGBoost achieve the highest accuracies, with XGBoost performing particularly better in the Arabic dataset. However, other papers used bigger datasets for their experiments, while ours was conducted on a smaller sample dataset, which could explain the differences in results. 

Our experiment was especially focused on demonstrating that adding OSINT features to a dataset increases a model's accuracy and overall performance. The results supported our hypothesis, except for one model, which had a slight decrease in accuracy. 

\section{Conclusion}
Our research experiment is aimed at testing the hypothesis that a model trained by an OSINT feature-enhanced dataset is more likely to outperform a model trained on a dataset without OSINT features. The machine learning classifiers implemented are DT, RF, SVM, XGBoost, and Multinomial NB. They were trained using our English and Arabic datasets, and the results obtained were evaluated using the following metrics: accuracy, f1 score, precision, recall, and confusion matrix. The models improved their accuracies after training on the OSINT-enhanced datasets, in both English and Arabic, with RF achieving the highest accuracy of 97.37\%.

Future research directions include using a larger dataset to include a diversified range of phishing and safe emails, as this would make our findings more generalized. Another direction is testing with deep learning transformers such as XLM-Roberta and GPT-3, which could be especially effective and suitable for our multilingual datasets.



\section*{Acknowledgment}

The authors would like to genuinely thank Professor Dr. Mehmet Nafiz Aydın of Kadir Has University for advising this research. His reviews and feedback have been invaluable, providing key motivation and direction in shaping the methodology, analysis, and overall quality of this work. 
\section*{Declaration of AI Tools Usage}

The authors of this study hereby declare that AI tools and technologies were utilized in the preparation of this research paper. Specifically, ChatGPT of OpenAI was employed as a coding assistant and for improving text readability, while Grammarly was used to enhance grammar correction and sentence structure. The use of these tools was limited to providing assistance in language and coding-related tasks, and the intellectual content, research ideas, and findings presented in this paper are solely the result of the authors’ work.

\ifCLASSOPTIONcaptionsoff
  \newpage
\fi

\bibliographystyle{IEEEtran}  
\bibliography{references.bib}

\begin{thebibliography}{10}
\providecommand{\url}[1]{#1}
\csname url@samestyle\endcsname
\providecommand{\newblock}{\relax}
\providecommand{\bibinfo}[2]{#2}
\providecommand{\BIBentrySTDinterwordspacing}{\spaceskip=0pt\relax}
\providecommand{\BIBentryALTinterwordstretchfactor}{4}
\providecommand{\BIBentryALTinterwordspacing}{\spaceskip=\fontdimen2\font plus
\BIBentryALTinterwordstretchfactor\fontdimen3\font minus \fontdimen4\font\relax}
\providecommand{\BIBforeignlanguage}[2]{{%
\expandafter\ifx\csname l@#1\endcsname\relax
\typeout{** WARNING: IEEEtran.bst: No hyphenation pattern has been}%
\typeout{** loaded for the language `#1'. Using the pattern for}%
\typeout{** the default language instead.}%
\else
\language=\csname l@#1\endcsname
\fi
#2}}
\providecommand{\BIBdecl}{\relax}
\BIBdecl

\bibitem{orunsolu2019predictive}
A.~Orunsolu, A.~S. Sodiya, and A.~T. Akinwale, ``A predictive model for phishing detection,'' \emph{Journal of King Saud University - Computer and Information Sciences}, vol. 232--247, 2019.

\bibitem{de2020impact}
R.~De, N.~Pandey, and A.~Pal, ``Impact of digital surge during covid-19 pandemic: a viewpoint on research and practice,'' \emph{International Journal of Information Management}, vol.~55, no. 102171, 2020.

\bibitem{shombot2024application}
E.~S. Shombot, G.~Dusserre, R.~Bestak, and N.~B. Ahmed, ``An application for predicting phishing attacks: A case of implementing a support vector machine learning model,'' \emph{Cyber Security and Applications}, vol.~2, p. 100036, 2024.

\bibitem{fette2007learning}
I.~Fette, N.~Sadeh, and A.~Tomasic, ``Learning to detect phishing emails,'' in \emph{Proceedings of the 16th International Conference on World Wide Web (WWW ’07)}, 2007, pp. 649--656.

\bibitem{nordine2019osint}
\BIBentryALTinterwordspacing
J.~Nordine, ``Osint framework,'' 2019. [Online]. Available: \url{https://osintframework.com}
\BIBentrySTDinterwordspacing

\bibitem{kaggle_phishing}
\BIBentryALTinterwordspacing
S.~Chakraborty, ``Phishing email detection,'' 2023. [Online]. Available: \url{https://www.kaggle.com/dsv/6090437}
\BIBentrySTDinterwordspacing

\bibitem{kaggle_paper1}
\BIBentryALTinterwordspacing
A.~S, P.~R. Nishant, S.~Baitha, and K.~D. Kumar, ``An ensemble classification model for phishing mail detection,'' \emph{Procedia Computer Science}, vol. 233, pp. 970--978, 2024, 5th International Conference on Innovative Data Communication Technologies and Application (ICIDCA 2024). [Online]. Available: \url{https://www.sciencedirect.com/science/article/pii/S187705092400646X}
\BIBentrySTDinterwordspacing

\bibitem{kaggle_paper2}
O.~C. Çetlenbik, R.~Gürfidan, and A.~Süzen, ``Classification of phishing attacks using the roberta model,'' 03 2024.

\bibitem{kaggle_paper3}
L.~Zhou, A.~Gaurav, V.~Arya, R.~Attar, S.~Bansal, and A.~Alhomoud, ``Enhancing phishing detection in semantic web systems using optimized deep learning models,'' \emph{International Journal on Semantic Web and Information Systems}, vol.~20, pp. 1--13, 01 2024.

\bibitem{sricharangundla_2023_a}
S.~C.~G. et~al., ``A feature extraction approach for the detection of phishing websites using machine learning,'' \emph{Journal of circuits, systems, and computers}, 08 2023.

\bibitem{maryam_2023}
U.~Maryam, ``Phishing attacks facilitated by open-source intelligence,'' 10 2023.

\bibitem{rastenis_2021}
\BIBentryALTinterwordspacing
J.~Rastenis, S.~Ramanauskaitė, I.~Suzdalev, K.~Tunaitytė, J.~Janulevičius, and A.~Čenys, ``Multi-language spam/phishing classification by email body text: Toward automated security incident investigation,'' \emph{Electronics}, vol.~10, no.~6, 2021. [Online]. Available: \url{https://www.mdpi.com/2079-9292/10/6/668}
\BIBentrySTDinterwordspacing

\bibitem{staples_2023}
D.~Staples, S.~Hakak, and P.~Cook, ``A comparison of machine learning algorithms for multilingual phishing detection,'' 08 2023, pp. 1--6.

\bibitem{bryushinin_2022}
A.~O. Bryushinin, A.~V. Dushkin, and M.~A. Melshiyan, ``Automation of the information collection process by osint methods for penetration testing during information security audit,'' in \emph{2022 Conference of Russian Young Researchers in Electrical and Electronic Engineering (ElConRus)}, 2022, pp. 242--246.

\bibitem{mosa_2023}
D.~Mosa, M.~Shams, A.~Abohany, E.-S. El-kenawy, and M.~Thabet, ``Machine learning techniques for detecting phishing url attacks,'' \emph{Computers, Materials \& Continua}, vol.~75, pp. 1271--1290, 01 2023.

\bibitem{emmanuel_2024}
\BIBentryALTinterwordspacing
E.~S. Shombot, G.~Dusserre, R.~Bestak, and N.~B. Ahmed, ``An application for predicting phishing attacks: A case of implementing a support vector machine learning model,'' \emph{Cyber Security and Applications}, vol.~2, p. 100036, 2024. [Online]. Available: \url{https://www.sciencedirect.com/science/article/pii/S277291842400002X}
\BIBentrySTDinterwordspacing

\bibitem{alnemari_2023}
\BIBentryALTinterwordspacing
S.~Alnemari and M.~Alshammari, ``Detecting phishing domains using machine learning,'' \emph{Applied Sciences}, vol.~13, no.~8, 2023. [Online]. Available: \url{https://www.mdpi.com/2076-3417/13/8/4649}
\BIBentrySTDinterwordspacing

\bibitem{fukan_2024_phishing}
F.~Çolhak Furkan~et al., ``Phishing website detection through multi-model analysis of html content,'' \emph{arXiv (Cornell University)}, 01 2024.

\bibitem{nmap_ip}
\BIBentryALTinterwordspacing
``Finding an organization's ip addresses | nmap network scanning,'' nmap.org. [Online]. Available: \url{https://nmap.org/book/host-discovery-find-ips.html}
\BIBentrySTDinterwordspacing

\bibitem{nmap_max}
\BIBentryALTinterwordspacing
``Timing and performance | nmap network scanning,'' Nmap.org. [Online]. Available: \url{https://nmap.org/book/man-performance.html}
\BIBentrySTDinterwordspacing

\bibitem{samrudhamhatre_2024_osintbased}
S.~Mhatre, F.~Schwarz, K.~Schwarz, and R.~Creutzburg, ``Osint-based email investigation,'' \emph{Electronic Imaging}, vol.~36, pp. 328--17, 01 2024.

\bibitem{feature-ext2}
J.~Mehanović~D., Kevrić, ``Phishing website detection using machine learning classifiers optimized by feature selection,'' \emph{International Information and Engineering Technology Association}, 2020.

\bibitem{feature-ext3}
R.~Mahajan and I.~Siddavatam, ``Phishing website detection using machine learning algorithms,'' \emph{International Journal of Computer Applications}, vol. 181, pp. 45--47, 10 2018.

\end{thebibliography}

\begin{IEEEbiography}
[{\includegraphics[width=1in, height=1.25in, clip, keepaspectratio]{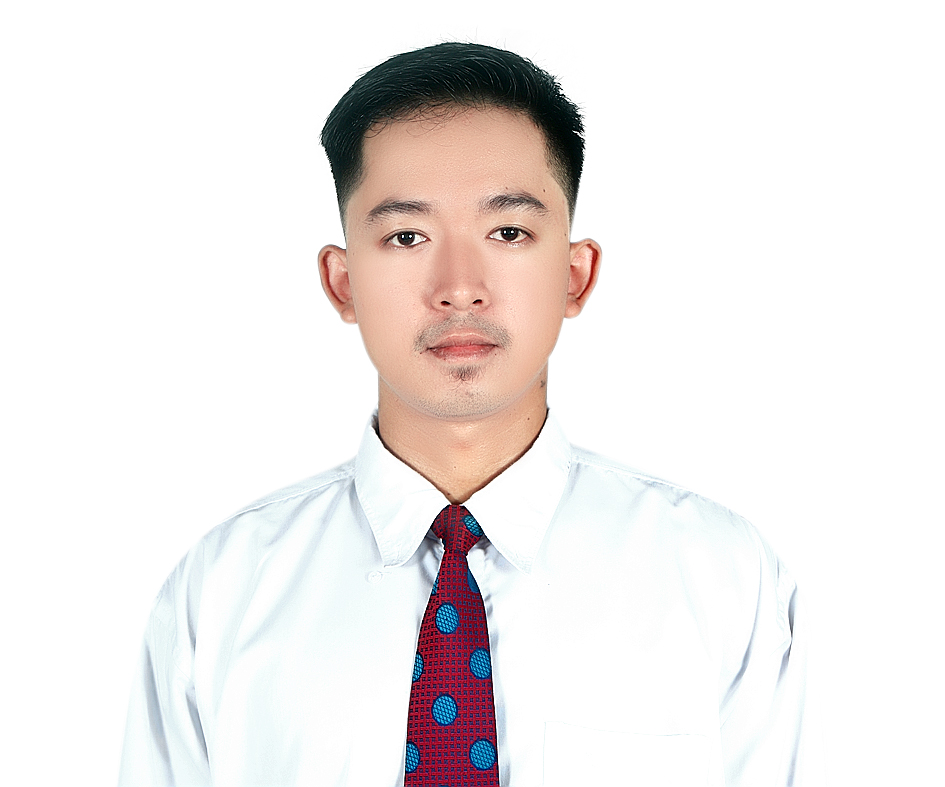}}]%
{Panharith An} earned his Telecommunication \& Electronic BEng in 2018 and spent five years as a full-stack web developer in the private and public sectors in Cambodia. His most recent role was a two-year tenure as Senior Software Engineer for the Digital Government Committee of Cambodia.
He is currently a fully-funded Year 1 Erasmus Mundus scholar in CyberMACS, a Joint Master’s Degree program in Applied Cybersecurity in Turkey and Germany, funded by the European Union. His current campus is Kadir Has University in Istanbul, Turkey.
\end{IEEEbiography}

\begin{IEEEbiography}
[{\includegraphics[width=1in, height=1.25in, clip, keepaspectratio]{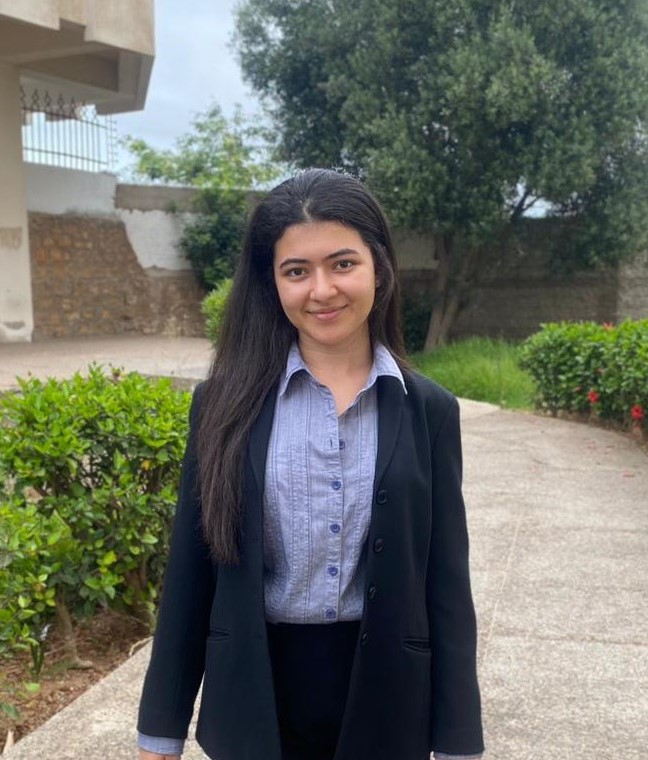}}]%
{Rana Shafi}
holds a Bachelor's degree in Mathematics and Computer Science in 2024. She is currently pursuing an Erasmus Mundus joint master's degree in Applied Cybersecurity in Turkey. Her research interests include Artificial Intelligence, Deep Learning, Computer Networking, and Network Security.
\end{IEEEbiography}

\begin{IEEEbiography}
[{\includegraphics[width=1in, height=1.25in, clip, keepaspectratio]{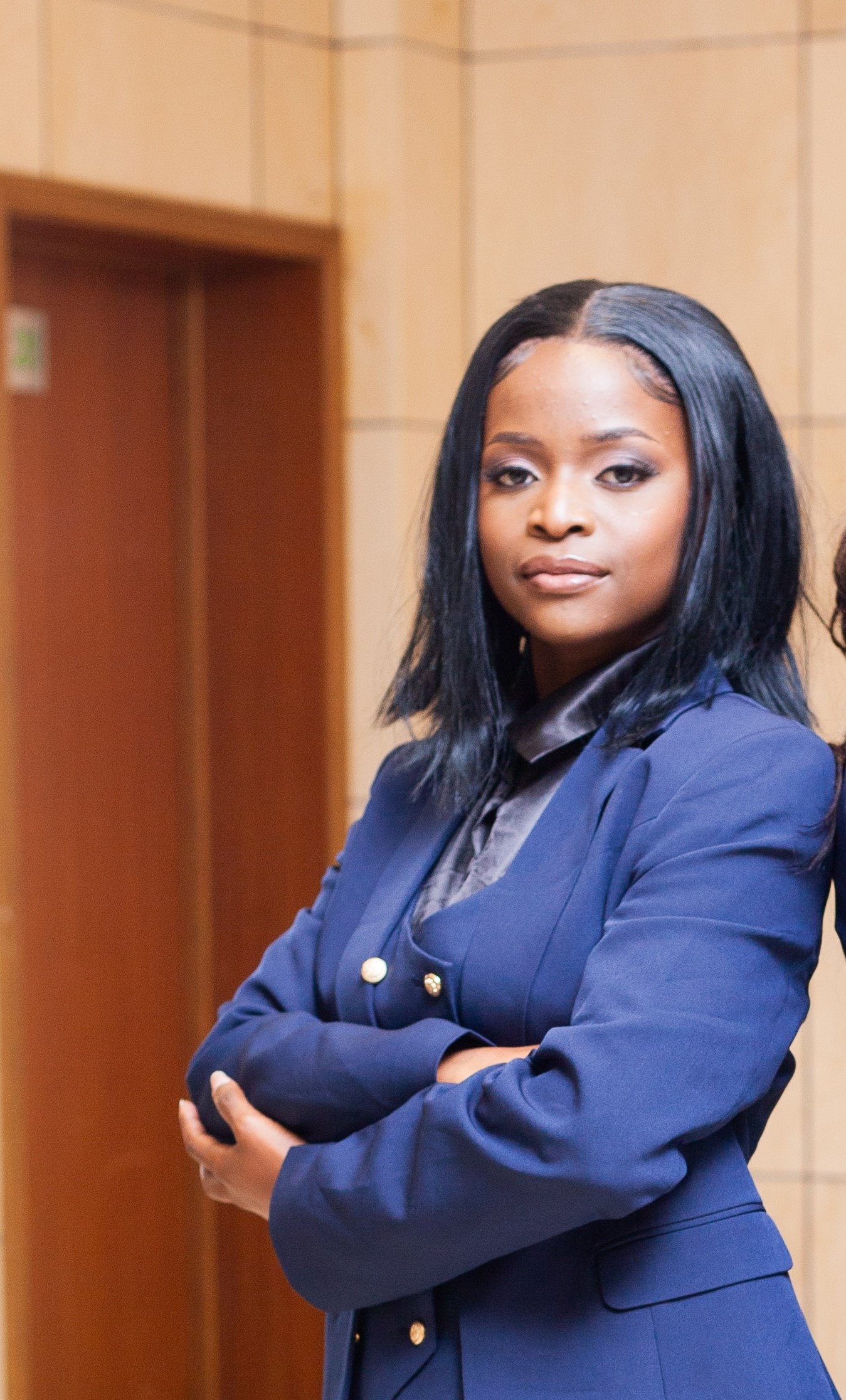}}]%
{Tionge Mughogho} holds a bachelor's degree in Computer Systems and Security. With two years of professional experience in software development and cybersecurity—spanning the national CERT of Malawi and the banking industry—they have honed their technical skills in critical areas. Tionge is currently pursuing a master's degree in Advanced Cybersecurity through the Erasmus Mundus scholarship at Kadir Has University in Istanbul, Turkey, furthering their expertise in the field.
\end{IEEEbiography}

\begin{IEEEbiography}
[{\includegraphics[width=1in, height=1.25in, clip, keepaspectratio]{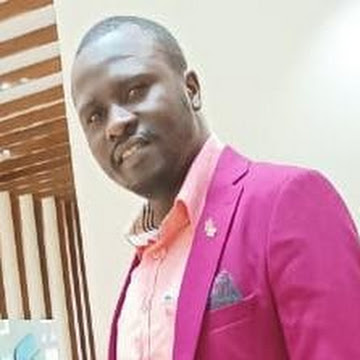}}]%
{Onyango Allan Onyango}
received his master's degree in Pure Mathematics in 2023. He has 4 years of professional experience in Computer Networks. He is presently pursuing an Erasmus Mundus joint master’s degree in Applied Cybersecurity at Kadir Has University, Turkey. His research interests include TDA, Network Security, Computer Networks, Operating Systems, and Cybersecurity Management.
\end{IEEEbiography}
\end{document}